# Sub 20 nm Short Channel Carbon Nanotube Transistors

*R. V. Seidel,\* A. P. Graham, J. Kretz, B. Rajasekharan, G. S. Duesberg, M. Liebau, E. Unger, F. Kreupl, and W. Hoenlein*

Infineon Technologies AG, Corporate Research, 81730 Munich, Germany

robert.seidel@infineon.com

*Corresponding author. Phone: (+49) 89 234-52755, E-mail: robert.seidel@infineon.com

Carbon nanotube field-effect transistors with sub 20 nm long channels and on/off current ratios of $> 10^6$ are demonstrated. Individual single-walled carbon nanotubes with diameters ranging from 0.7 nm to 1.1 nm grown from structured catalytic islands using chemical vapor deposition at 700 °C form the channels. Electron beam lithography and a combination of HSQ, calix[6]arene and PMMA e-beam resists were used to structure the short channels and source and drain regions. The nanotube transistors display on-currents in excess of 15 µA for drain-source biases of only 0.4 V.

The exceptional properties of carbon nanotubes (CNTs), including ballistic transport and semiconducting behavior with band-gaps in the range of 1 eV, have sparked a large number of theoretical[1-3] and experimental[4-6] studies. The possibility to use CNTs to replace crystalline silicon as the basis for high-performance transistors has prompted significant effort to reduce the size of CNT field-effect transistors (CNTFETs) in an effort to understand their scaling behavior and ultimate limits. In this context we present CNT transistors with channel lengths below 20 nm, which have characteristics comparable to those of much larger silicon-based demonstrators with similar channel lengths.[7,8]

Since the first CNTFET was demonstrated in 1998,[9] their characteristics have been continuously and rapidly improved, particularly in the last few years.[5,6] One critical aspect is the optimization of the source and drain contacts to minimize the Schottky barrier (SB) due to the mismatch between the CNT and contact metal work functions. As discussed by Guo et al.[1] the formation of significant SBs of the order of 0.2 eV can severely affect the function of short channel CNTFETs because electron injection through the SBs at higher drain-source biases reduces the on/off current ratio. In addition, substantial SBs reduces the on-current, as shown by the difference between Ti and Pd contacted CNTFETs.[6,5]

A further important aspect is the band gap of the CNTs which scale as $0.9/d$ eV, where $d$ is the nanotube diameter in nanometers.[10] As shown by Javey et al.[11] it is possible to form almost perfect ohmically contacted CNTFETs with 1.5-2 nm diameter nanotubes using Pd contacts. These devices have on-conductivities close to the quantum conduction limit $2G_0 = 155$ µS. However, these CNTs have band gaps of only 0.4-0.5 eV, leading to high off-currents for short channel devices.[5] Therefore, we have concentrated on CNTs with smaller diameters of around 0.7–1.1 nm that have larger band gaps of around 0.8-1.3 eV and are suitable for short channel devices.[12]

Catalytic chemical vapor deposition (CCVD) provides the possibility to selectively and cleanly grow CNTs with a narrow diameter distribution for the production of significant numbers of device demonstrators[12] or high current transistors.[13] In a recent publication we have demonstrated the possibility to grow small diameter CNTs for long channel CNTFETs gated using the substrate[12] or e-beam defined top gates.[14] In this work we have investigated the scalability of these devices to channel lengths below 20 nm.

The single-walled carbon nanotubes (SWCNTs) were selectively grown on highly doped n-type silicon substrates covered with a 12 nm thick thermal oxide dielectric. Catalyst islands with a separation of 2 µm consisting of a 10 nm thick diffusion barrier, 10 nm Al catalyst support layer and the 0.2 nm thick Ni catalyst layer were produced with optical lithography and lift-off using acetone. SWCNT growth was carried out at 700 °C in a quartz tube oven using methane following a preconditioning step in hydrogen, as described in Ref. 12. Tight control of the Ni catalyst thickness ensured that, on average, only one SWCNT bridged the 2 µm gap between the catalyst islands. The exact number of CNTs forming the channel was determined using high-resolution scanning electron microscopy (SEM) following measurement of the transistor characteristics. Atomic force microscopy (AFM) investigations of the CVD CNTs show that their diameters range from 0.7 nm to 1.1 nm.[12] The diameters of the CNTs in each transistor could not be determined due to the AFM tip curvature, which prevents measurements in the narrow gap between source and drain electrodes. Transmission electron microscopy images of CNTs grown under identical conditions and Raman spectroscopy confirm that the nanotubes are SWCNTs with diameters compatible with the AFM results.[12]

Electron beam (e-beam) lithography (FEI Altura DB835) at 30 kV in combination with 60 nm thick hydrogen-silsesquioxane (HSQ) or calix[6]arene negative-tone resists was used to create 10-25 nm wide $SiO_x$ or calix[6]arene lines between the catalyst pads, defining the channel length. Subsequent e-beam exposure of an additional 140 nm thick polymethylmethacrylate (PMMA) positive tone resist at 10 kV and lift-off in acetone were used to align the 10-12 nm thick palladium source and drain contacts on either side of the channel, as described in the process flow shown in Figure 1. A long-throw (vertical) e-beam evaporation system was used to deposit the thin Pd layers which resulted in slightly longer channels due to shadowing by the fin as the deposition progressed. The HSQ lines could be subsequently removed using a 100:1 buffered hydrofluoric acid solution. SEM images of the channel regions of CNTFETs prior and after the HSQ line removal are shown in Figure 2 and Figure 3, respectively. The highly doped silicon substrate was used as a back-gate to control the conduction in the short CNT channel.

The gate voltage ($V_{gs}$) dependence of the source-drain current ($I_{sd}$) of an 18 nm channel CNTFET is presented in Figure 4. The on/off current ratio is in excess of $10^6$ although higher on-currents were observed at higher drain-source bias ($V_{ds}$). No leakage from the source or drain contacts to the gate electrode through the thin oxide was observed (measurement limit 10 fA). The on conductance of 11 µS (1.1 µA at 0.1 V bias) corresponds to a normalized conductance of 12000 µS/µm, taking the nanotube diameter as the channel width. The transconductance amounts to 3.5 µS per nanotube or 4000 µS/µm. Measurements of the maximum current possible through these short channels was found to be above 100 µA, or a current density of $> 10^{10}$ A/cm$^2$, for metallic and small band gap CNTs and in excess of 15 µA, or $> 10^9$ A/cm$^2$, for semiconducting CNTs, compatible with ballistic conduction. However, the main current limitation for the semiconducting tubes appears to be the high applied fields, which destroy the device by impact ionization whereas heating at the contacts of the metallic SWCNTs leads to physical modification and damage of the contacts, which are made of only 10-12 nm thick Pd.

Transfer characteristic of a different SWCNT FET with 20 nm channel length are plotted in Figure 5a. The transfer characteristics were measured at $V_{ds}$ = 100 mV resulting in an on-current of as much as 4 µA. The corresponding output characteristic of the same nanotube is plotted in Figure 5b. The output characteristics could not be measured up to current saturation since long application of high currents would modify the contacts by overheating of the thin Pd metallization. The short channel nanotube transistors showed rather high noise in the output characteristics when repeatedly measured. Possible reasons are local heating at the contacts as well as the lack of a passivation which makes the transistors very sensitive to the ambient conditions.

As shown in Figure 6, more current can be transported by the CNTs at higher source-drain biases, in this third example up to 15.5 µA at $V_{ds}$ = 0.4 Volt. This corresponds to a conductivity of 38.8 µS or 0.5 $G_0$, suggesting that there is very little loss at the contacts, i.e. the Schottky barriers are small. A further important aspect is the absence of ambipolar behavior that is observed for Schottky barrier and smaller band-gap CNTFETs. We suggest that this is due to the low source-drain bias of $V_{ds}$ = 0.4 V relative to the band-gap of the narrow diameter ($d \sim 0.7 – 1.1$ nm, gap $\sim 0.8 - 1.3$ eV). The origin of the relatively high off-current of this device of around 43 pA is not clear, but the off-current is very sensitive to the humidity in the measurement environment. By purging the measurement apparatus with dry air the off-currents in all measured devices could be significantly reduced. For a number of 30 investigated short channel nanotube transistors (length ~ 18-20 nm), the on-currents ranged between 1 – 4 µA at $V_{ds}$ = 100 mV and 4 – 15 µA at $V_{ds}$ = 400 mV.

The relatively high subthreshold slopes of 170-200 mV/decade, and corresponding high $I_{off}$ at $V_{gs}$ = 0 V, are a result of the 12 nm thick $SiO_2$ gate oxide which, coupled with the short 18 nm source-drain separation, leads to poor gate-channel coupling. This can be clearly seen when the source-drain separation is increased to 360 nm while maintaining the same gate oxide thickness and composition, as shown in Figure 7. In this case the source and drain contacts were directly patterned using PMMA without the intermediate HSQ or calix[6]arene steps. The subthreshold slope for this device is substantially lower at about 80 mV/decade. Similar results were also obtained using thin, high dielectric constant (high-$k$) gate oxide materials. 70 mV/decade for a 3 µm long top-gate devices using an 8 nm thick $ZrO_2$ ($k$~25) gate oxide has been reported,[15] which is close to the room temperature theoretical minimum of 60 mV/decade. Therefore, we anticipate that better switching behavior will be obtained for the current devices with thinner, high-$k$ gate oxides. In addition, it is anticipated that the parallel operation of a number of CNTs will lead to a significant performance improvement of individual devices.[13]

In conclusion, we have demonstrated the fabrication and operation of sub 20 nm channel single-walled carbon nanotube field-effect transistors. To our best knowledge, the ~18 nm channel length CNT transistor represents the shortest operating nanotube transistor fabricated to date. Without significant optimization, these transistors already perform as well as much larger state-of-the-art silicon-based devices in many aspects. The high on/off current ratio indicates that further scaling is possible.

The authors would like to acknowledge technical support from W. Weber and B. Panzer. This project is supported by BMBF under the INKONAMI project 13N8402.

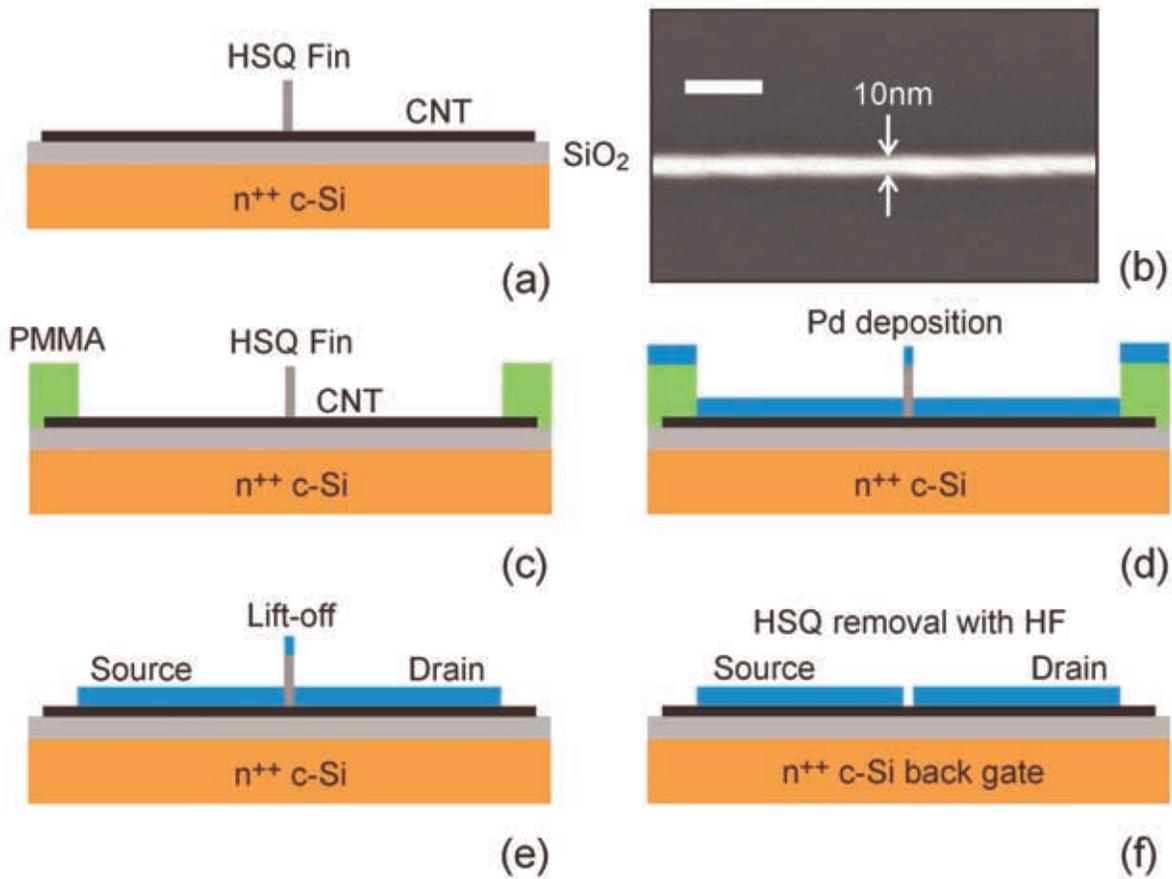

**Figure 1.** Process flow for the creation of short channel CNTFETs. (a) The channel is defined with e-beam using HSQ negative tone resist on top of the CVD grown SWCNTs. An SEM image of a 10 nm wide HSQ line is shown in (b) where the scale bar is 50 nm. This is followed by source and drain region structuring with PMMA positive tone resist (c). Source and drain contacts are formed by the deposition of 10-12 nm Pd (d) and lift-off in acetone (e). The channel is freed using a buffered HF wet-etch (f). The highly doped Si substrate serves as the gate.

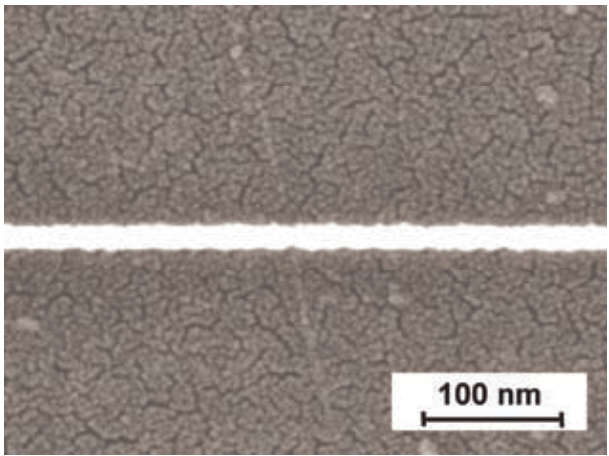

**Figure 2.** SEM of a short channel CNTFET after metal deposition and prior to HSQ fin removal. The CNT is just about visible under the 10 nm thick Pd layer, whereas the HSQ fin with the metal on top is very prominent. The fin width is only 15 nm but, because the CNT does not lie perpendicular to the fin and due to shadowing during the metal deposition, the effective channel length is higher (18–20 nm).

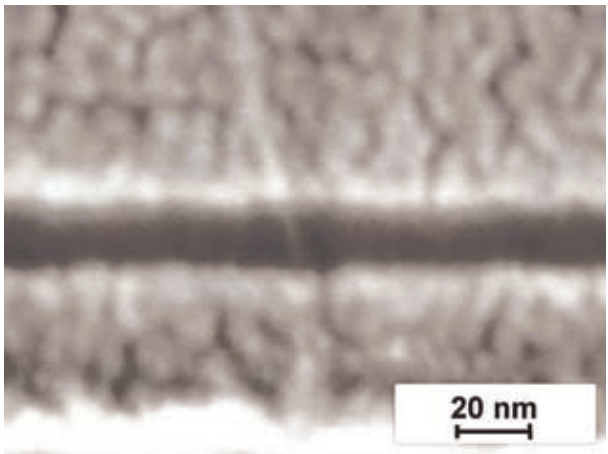

**Figure 3.** SEM image of the source and drain contacts and the individual carbon nanotube forming a field-effect transistor. The exposed SWCNT length between the source and drain contacts is about 18 nm. The HSQ fin covering the channel region has been removed using buffered fluoric acid to expose the CNT.

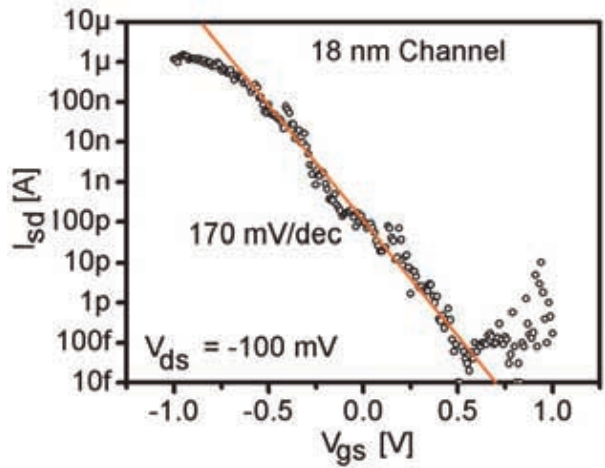

**Figure 4.** Gate voltage dependence of the source-drain current ($I_{sd}$) in an ~18 nm channel CNTFET for a drain-source voltage of $V_{ds}$ = -0.1 V. The as-grown SWCNTs show p-type behavior. The current varies between $I_{sd} \approx$ 100 fA in the off-state at $V_{gs}$ = +0.5 V and $I_{sd}$ = 1.1 µA in the on-state at $V_{gs}$ = -1.0 V, i.e. seven orders of magnitude. The subthreshold slope is 170 mV/decade and the transconductance is 3.5 µS (4000 µS/µm).

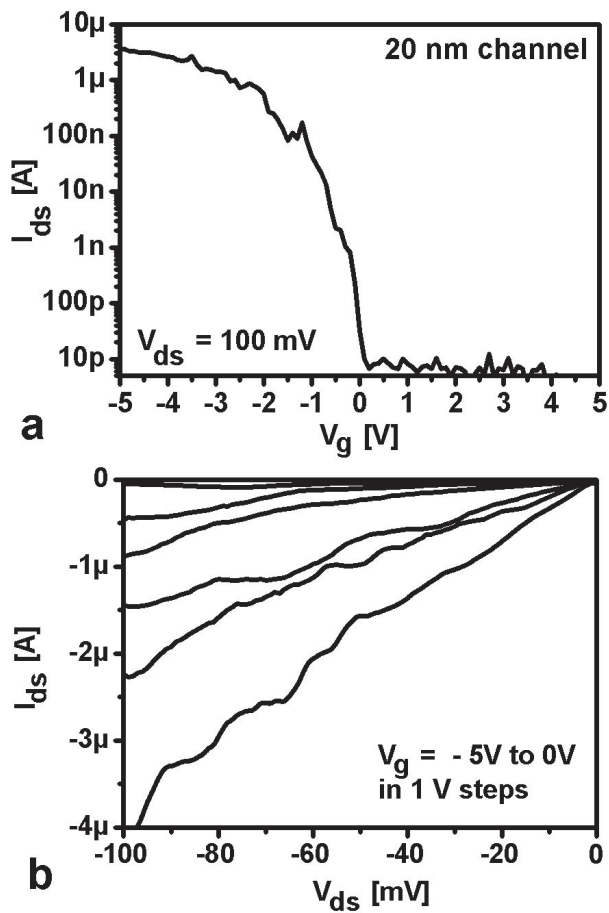

**Figure 5.** Electrical properties of a ~20 nm long nanotube transistor. (a) Transfer characteristic taken at $V_{ds}$ = 100 mV. The lower bound current measurement of this device was limited to 10 pA by the used measurement instrument and the off-current could even be lower. (b) Output characteristics of the same device. The gate voltage was increased from $V_g$ = -5 V to 0 V in 1 V steps reducing the output current of the p-type nanotube with every sweep of $V_{ds}$.

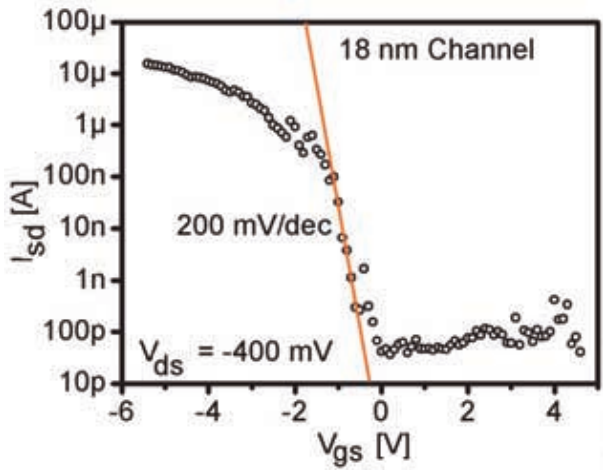

**Figure 6.** Gate voltage dependence of the source-drain current in an ~18 nm channel CNTFET for a drain-source voltage of $V_{ds}$ = -0.4 V. The as-grown SWCNTs show p-type behavior. The current varies between $I_{sd} \approx 43$ pA in the off-state at $V_{gs}$ = 0 V and $I_{sd}$ = 15.5 µA in the on-state at $V_{gs}$ = -5.4 V. The subthreshold slope is ~200 mV/decade and the transconductance is 6.75 µS.

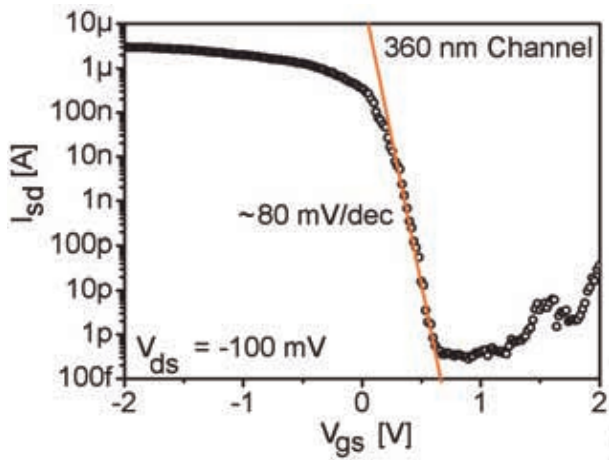

**Figure 7.** Gate voltage characteristics for a 360 nm long channel CNTFET with a 12 nm thick $SiO_2$ gate oxide. The subthreshold slope of around 80 mV/decade is much lower than for the short channel devices due to the improved electrostatic geometry. The transconductance of this device is 1.85 µS.